\documentclass[12pt,preprint]{aastex}

\usepackage{graphicx}

\newcommand\teff{T$_{\rm{eff}}$}

\newcommand\msini{$M_p$ sin\textit{i}}

\newcommand\earthmass{$M_{\oplus}$}

\newcommand\solmass{$M_{\odot}$}

\begin{document}

\title{Forming Gas Giants Around a Range of Protostellar M-dwarfs by Gas Disk Gravitational Instability}

\author{Alan P. Boss \& Shubham Kanodia}
\affil{Earth \& Planets Laboratory, Carnegie Institution
for Science, 5241 Broad Branch Road, NW, Washington, DC 20015-1305}
\authoremail{aboss@carnegiescience.edu}

\begin{abstract}
Recent discoveries of gas giant exoplanets around M-dwarfs (GEMS) from transiting and radial velocity (RV) surveys are difficult to explain with core-accretion models. We present here a homogeneous suite of 162 models of gravitationally unstable gaseous disks. These models represent an existence proof for gas giants more massive than 0.1 Jupiter masses to form by the gas disk gravitational instability (GDGI) mechanism around M-dwarfs for comparison with observed exoplanet demographics and protoplanetary disk mass estimates for M-dwarf stars. We use the Enzo 2.6 adaptive mesh refinement (AMR) 3D hydrodynamics code to follow the formation and initial orbital evolution of gas giant protoplanets in gravitationally unstable gaseous disks in orbit around M-dwarfs with stellar masses ranging from 0.1 $M_\odot$ to 0.5 $M_\odot$. The gas disk masses are varied over a range from disks that are too low in mass to form gas giants rapidly to those where numerous
gas giants are formed, therefore revealing the critical disk mass necessary for gas giants to form by the GDGI mechanism around M-dwarfs. The disk masses vary from 0.01 $M_\odot$ to 0.05 $M_\odot$ while the disk to star mass ratios explored range from 0.04 to 0.3. The models have varied initial outer disk temperatures (10 K to 60 K) and varied levels of AMR grid spatial resolution, producing a sample of expected gas giant protoplanets for each star mass. Broadly speaking, disk masses of at least 0.02 $M_\odot$ are needed for the GDGI mechanism to form gas giant protoplanets around M-dwarfs.
\end{abstract}

% The resulting homogeneous suite of 162 Enzo 2.6 models represents an attempt at population synthesis models for gas giant exoplanets more massive than 0.1 Jupiter masses, suitable for comparison with observed exoplanet demographics and protoplanetary disk mass estimates for M-dwarf stars. 


\section{Introduction}

 The Kepler Mission provided the first demographic census of exoplanets transiting solar-type (FGK) stars (Borucki et al. 2010; Bryson et al. 2021). Combined with the results of ground-based Doppler spectroscopic studies of solar-type stars (e.g., Butler et al. 2017), we now have a reasonably complete demographic census for exoplanets orbiting solar-type stars, at least on relatively short period orbits. These observational constraints have fueled the creation of theoretical population synthesis models, which attempt to use the predictions of detailed simulations of planet formation by the classic core accretion mechanism (e.g., Ida et al. 2018) or the heretical gas disk gravitational instability (GDGI) mechanism (e.g., Boss 2023) to match the demographics for solar-type stars. 

M-dwarfs (\teff{} $<$ 4000 K) are of particular interest in the search for habitable worlds and biosignatures (e.g., Scalo et al. 2007; Childs et al. 2022; Peterson et al. 2023),  as they outnumber solar-type stars by an order of magnitude and the compactness of their habitable zones makes them prime targets for characterization of exoplanets that transit and are eclipsed by their host star (e.g., TRAPPIST-1, Greene et al. 2023). M-dwarfs constituted only a small fraction of the target stars for the Kepler Mission, yet due to the prevalence of small terrestrial planets around these stars we now have robust estimates on the occurrence of these planets (Dressing \& Charbonneau 2015; Hardegree-Ullman et al. 2019; Hsu et al. 2020).  However there is still much to be learned observationally about the demographics of giant exoplanets around M-dwarf stars (GEMS). Understanding giant planet demographics is not just crucial to get a holistic picture of planet formation, but also because giant exoplanets are known to influence the formation and evolution of terrestrial planets (Childs et al. 2019; Mulders et al. 2021; Schlecker et al. 2021; Bitsch and Izidoro 2023).

Over the last decade, Doppler surveys are increasingly extending their target lists to include M-dwarfs (Bonfils et al. 2013; Feng et al. 2020) or are exclusively targetting M-dwarfs (Mahadevan et al. 2014; Kotani et al. 2018; Sabotta et al. 2021). M-dwarf exoplanets have been found by Doppler spectroscopy with minimum masses ranging from Earth-mass (e.g., Kossakowski et al. 2023), to Neptune-mass (e.g., Blanco-Pozo et al. 2023, Stefansson et al. 2023), to Jupiter-mass (e.g., Endl et al. 2022), to super-Jupiter-mass (e.g., Sozzetti 2023).  In comparison to the constrained Kepler fields of view, the Transiting Exoplanet Survey Satellite (TESS) mission targets stars distributed across the entire sky, and hence can detect transiting exoplanets orbiting relatively bright nearby M-dwarfs (Muirhead et al. 2018). This has been helpful to detect and confirm a number of giant exoplanets around M-dwarf stars (GEMS; Johnson et al. 2012; Canas et al. 2020; Canas et al. 2022; Jordan et al. 2022; Kanodia et al. 2022; Lin et al. 2023; Canas et al. 2023; Kanodia et al. 2023; Hobson et al. 2023; Kagetani et al. 2023). These detections from Doppler and transiting survey are a challenge to explain for core-accretion (Morales et al. 2019; Schlecker et al. 2022; Kanodia et al. 2023).

% Susemiehl \& Meyer (2022) estimated that the binary fraction of M-dwarfs is between 0.229 and 0.462, depending on the binary mass ratio ($q$) that is assumed, with the larger fraction applying to $q \ge 0.1$ for the latter, compared to $q \ge 0.6$ for the former estimate.

Here we will explore the predictions of the gas disk gravitational instability (GDGI) mechanism (e.g., Boss 1997) for gas giant protoplanet formation for the demographics of exoplanets orbiting M-dwarfs with a range of masses, $0.1~M_\odot$ to $0.5~M_\odot$. Most of the first author's work on the GDGI mechanism has focused on solar-mass protostars (e.g., Boss 2023), with an emphasis on improving the spatial resolution of the 3D hydrodynamics code models (e.g., Boss 2021a) or the treatment of the thermodynamical evolution of the disk (e.g., Boss 2021b). While several groups have studied the GDGI for M dwarf primaries (Mercer \& Stamatellos 2020; Haworth et al. 2020),
only a handful of the first author's GDGI models have dealt with M-dwarf primaries (Boss 2006; Boss 2011), not enough to make a meaningful prediction for population synthesis models. Our present suite of models is intended to address this shortfall in GDGI models for M-dwarf stars. Note, however, that our work is only a first step toward a true population synthesis model for GDGI, as we do not try to vary, e.g., the assumed protoplanetary disk radii, as was done by Boss (2023), or to predict the final orbital parameters of the resulting gas giant protoplanets. We leave these key steps to future work.

 Our models of protoplanet formation and early orbital evolution employ the Enzo 2.6 hydrodynamics code and follow from the calculations performed by Boss (2023) using Enzo 2.5 for solar-mass protostars. As noted by Boss (2023), Enzo is a three dimensional (3D) hydrodynamic code that uses Adaptive Mesh Refinement (AMR) in Cartesian coordinates to ensure that sharp gradients in fluid quantities such as shock fronts can be handled accurately. Enzo is able to replace exceptionally dense disk clumps with sink particles representing newly formed, self-gravitating protoplanets, which thereafter interact gravitationally with each other and with the disk while accreting disk gas, growing in mass to the gas giant planet range and possibly beyond.
 
\section{Numerical Hydrodynamics Code}

The numerical calculations were performed with the Enzo AMR hydrodynamics code (Colella \& Woodward 1984; Collins et al. 2010; Turk et al. 2011; Bryan et al. 2014). Boss (2023) used the Enzo code to perform a similar set of calculations, and can be consulted for the details of the implementation. Here we limit ourselves to noting explicitly any differences between the Boss (2023) calculations and the present work. Boss (2023) used Enzo 2.5, whereas the present work used Enzo 2.6, which should have no effect on the calculations. A maximum of 6 levels of refinement was used, with a factor of two refinement occurring for each level, so that the maximum possible effective grid resolution was $2^6$ = 64 times higher than the initial top grid resolution of $32^3$, i.e., $2048^3$. The cubic computational box had sides 120 au in length. A point source of external gravity was used to represent an accreting protostar of varied mass at the center of the grid. Newly formed gas giant protoplanets are represented by sink particles in the models (Wang et al. 2010). Sink particle creation was only allowed for cells with densities exceeding $10^{-9}$ g cm$^{-3}$. As in Boss (2023), radiative cooling in optically thin regions was employed in the models. The models do not include heating by the central protostar, given the high optical depth in the disk midplane, but do include the heating produced by compression of the disk gas in converging regions of fluid flow, e.g., spiral arms.

\section{Initial Conditions}

The Enzo 2.6 initial disk models are based on the model HR disk from Boss (2001), which has been used extensively in the first author's GDGI models (e.g., Boss 2017, 2021a, 2021b, 2023). The underlying initial density distribution is not an assumed power-law surface density profile, but rather is that of an axisymmetric, two-dimensional, adiabatic, self-gravitating, thick disk in near-Keplerian rotation about a stellar mass $M_s$ (Boss 1993):

$$ \rho(R,Z)^{\gamma-1} = \rho_o(R)^{\gamma-1} - \biggl( 
{ \gamma - 1 \over \gamma } \biggr) \biggl[
\biggl( { 2 \pi G \sigma(R) \over K } \biggr) Z +
{ G M_s \over K } \biggl( { 1 \over R } - { 1 \over (R^2 + Z^2)^{1/2} }
\biggr ) \biggr], $$

\noindent where $R$ and $Z$ are cylindrical coordinates and $\sigma(R)$ is 
a nominal surface density. The adiabatic constant $K = 1.7 \times 10^{17}$ (cgs units) and adiabatic exponent $\gamma = 5/3$ are chosen based on the thermodynamics of collapsing molecular clouds (Boss 1993). The initial midplane density $\rho_o(R)$ is chosen to ensure near-Keplerian rotation throughout the disk:

$$\rho_o(R) = \rho_1 \biggl( {R_1 \over R} \biggr)^{3/2}, $$

\noindent 
where $R_1 = 1$ au, the inner edge of the numerical grid, and $\rho_1$ is varied in order to produce disks with different total masses, depending on the mass of the central protostar. The nominal surface density is also chosen to be a power-law in cylindrical radius $R$, $\sigma(R) \propto R^{-1/2}$. Figure 1 of Boss (2001) shows that this initial disk model leads to an initial surface density profile in spherical coordinate $r$ which is not a pure power-law profile, but is similar to a power-law surface density profile proportional to $r^{-1}$.

The initial disks have varied outer disk temperatures, ranging from 10 K to 60 K, and have disk envelope temperatures of 35 K. With an outer disk temperature of 40 K, model HR has an initial minimum Toomre $Q \approx 1.3$, implying marginal stability to the growth of rings and spiral arms (Toomre 1964). The initial disks have inner radii of 1 au  and outer radii of 40 au. A 120 au box size was used with these initially 40 au radius disks, allowing them to evolve and expand outward to 60 au before hitting the outer boundary. The disks orbit a central protostar with a mass varied from 0.1 $M_\odot$ to 0.5 $M_\odot$. The disk masses were varied from  0.01 $M_\odot$ to 0.05 $M_\odot$. The same disk density power-law-like Keplerian structure (Boss 2001) was used for all of the models, with the structure being terminated at 40 au. Figure 1 shows the cross sections of the initial disk density distributions, both parallel and perpendicular (i.e., in the disk midplane) to the disk rotation axis, for model M1-6-40-40 (see Table 1). The initial disk temperatures for all models are 1500 K inside 1 au, and fall linearly to the specified outer disk temperatures (see Tables) at 6 au.

The large disk masses explored in the models should be considered as possibly representative of the transition from massive protostellar disks to less massive protoplanetary disks, i.e., of Class 0 protostar disks. Whether such large masses are appropriate to consider for M-dwarf disks can only be determined by observational estimates of the disk (and host star) masses for the youngest protostellar objects. In Figure 6 we include a histogram of measured disk masses for protostellar Class 0/I disks from a VLA survey of $\sim$ 100 disks in Perseus (Table 10 from Tychoniec et al. 2018), for comparison with the assumptions from our simulations. Here we note the caveat that this survey was not focussed on M-dwarf disks, and therefore likely includes more massive disks (on average) than those around M-dwarfs. However, given that the disks that are prone to gravitational instability and forming GEMS are likely more massive than the median disk (and hence likely outliers), this comparison depicts the existence of such disks in existing samples.

\section{Results}

Tables 1 through 5 list the models for each mass protostar, with variations in the number of levels of grid refinement, the initial outer disk temperatures, initial disk masses, and finally the masses of the gas giant protoplanets (i.e., Enzo sink particles) that formed and survived to the final time of 100 yr of evolution. Sinks were not allowed to form for the first 10 yr of evolution in order to allow the initial disks to relax and begin expanding as a result of the rapid formation of spiral arms and their associated mass and angular momentum transport (e.g., Boss 1984). Sinks with masses less than 0.1 $M_{Jup}$ after 100 yr of evolution are not reported in the Tables; these formed occasionally in the models.

Of the 162 models shown in the Tables, 96 were run with a maximum of 6 levels of refinement, 63 with a maximum of 5 levels, and 3 with a maximum of 4 levels. The 4 levels models ran quickly (in hours) and produced results not too dissimilar from the 5 levels models (Table 5), but were considered too coarse in resolution to produce a reasonably reliable sink particle mass estimate. The 5 levels models ran faster than the 6 levels models (in days compared to weeks) and produced results not dissimilar from the 6 levels models, which can be considered the closest to approaching the continuum limit of resolution. Boss (2023) ran with Enzo AMR levels ranging from 3 to 7, but encountered small time step problems with the 7 levels model, and hence 7 level models were not considered in the present study. In lieu of the sort of Monte Carlo sampling that can be used for the chaotic orbital dynamics underlying core accretion models, the somewhat chaotic outcomes of running the present models with 5 or 6 levels should provide some measure of the inherent uncertainties in modeling the GDGI mechanism.

Given the fact that the initial disks are marginally gravitationally unstable, the initial evolution is rapid, with spiral arms and rings forming within the first 20 yr of evolution, concurrent with the formation of the first sink particles. In the case of the model shown in Figure 2, M1-6-40-40, 11 sink particles form by 20 yr and then grow in mass by merging and gas accretion until there are only two sink particles with masses greater than 0.1 $M_{Jup}$ by 100 yr and one sink particle with a mass less than 0.1 $M_{Jup}$, as seen in Figure 2. The two gas giants in Figure 2 are massive enough (2.5 $M_{Jup}$ and 5.5 $M_{Jup}$) to be encircled by their own accretion disks, fed by streamers of disk gas tied to the spiral arms and rings in the inner disk. Note the absence of large-scale, distinct one-armed spirals; the disk's spiral arms are more subtle in structure and appear more as segments of rings than as grand spirals in an extragalactic context. Note that Figure 2 shows only the innermost 6 au in radius of the 60 au half-box-size computational volume. This is because all the GDGI action occurs within 6 au in these models; the disk does not expand significantly beyond its initial orbital radius of 40 au by 100 yr. Similarly, all the action with the sink particles occurs within the inner disk, and none were ejected outward to exit the computational volume, unlike the case of the vigorous GDGI models with a solar-mass primary of Boss (2023).

Figure 2 also depicts the midplane temperature distribution at the end of the calculation for model M1-6-40-40, showing significant midplane temperature variations associated with compressional heating of the disk gas in dense regions coupled with radiative cooling to the surface of the disk, which can be seen in the cross-sections. The region shown in this Figure is the inner disk region which started with initial temperatures between 1500 K and the initial outer disk temperature of 40 K, showing that while still warmer than the initial outer disk, this region has cooled radiatively substantially from the initially imposed midplane temperature profile. A close examination reveals that some low density regions of the disk midplane are warmer than might have been expected for radiation to free space at $\sim$ 10 K; this appears to be due to radiative heating from the hotter regions of the cloud envelope, where the cloud envelope has heated up to over 100 K from the initial envelope temperature of 35 K. The Figure also shows that the initial disk (Figure 1) has contracted vertically considerably as a result of cooling, reaching midplane spatial densities nearly 100 times higher than in the initial disk model. 

We note that in model M1-6-40-40 the gas giants formed and orbited stably inside a few au, as is the case with all of the exoplanets formed in the 162 models presented here. Clearly this is a result of the particular initial conditions chosen for these disk models; other choices can produce wider orbit exoplanets (e.g., Boss 2011). Given that our goal here is primarily to synthesize populations of gas giants that can be compared with observational detections by transits and Doppler spectroscopy, not direct imaging, these models appear to offer an appropriate approach for predicting the likelihood of gas giant planet formation by GDGI that results in relatively short orbital periods.

The five Tables give the masses of the gas giant protoplanets that formed in all of the models and survived orbital migration to 100 yr of evolution. The disk masses for each stellar mass were chosen to span the boundary between disks too low in mass to undergo a successful GDGI to those so massive as to be assured of a robust GDGI. For each stellar mass, this is evidenced by the change in exoplanet masses as the disk masses are decreased: starting from multiple, massive exoplanets at the high end, the numbers and masses of the exoplanets decrease until none at all are formed. In each case, the most dramatic changes in outcome can be seen to be a result of the disk masses, with the initial outer disk temperatures playing a significant but more minor role. This latter effect appears to be a result of the approximate treatment of disk thermodynamics in the present models, with a crude treatment of radiative cooling and disk optical thickness, compared to GDGI models where a full treatment of radiative transfer is employed (e.g., Boss 2021b) or the $\beta$ cooling formalism (Gammie 2001; Boss 2021a), where the outcome can be considerably more dependent on the initial outer disk temperature and other assumptions about the thermodynamical treatment. However, three dimensional radiative hydrodynamics models can be notoriously slow to compute; each such model in Boss (2021b) required about 4.5 yr of processing on a dedicated cluster core. The present models are intended to provide a reasonably accurate estimate of the dependence of the GDGI mechanism on stellar and disk mass, in order to yield a zeroth order comparison with observations of M-dwarf disk masses and M-dwarf gas giant exoplanet demographics.

Figures 3, 4, and 5 present the results from the five Tables in a graphical format. Figure 3 shows the dependence of gas giant protoplanet formation by GDGI on the disk mass and the initial outer disk temperature, for each of the five M-dwarf masses investigated. A strong dependence on disk mass is apparent. Figure 4 presents the final exoplanet masses for the 162 models as a function of disk mass for each M-dwarf mass, clearly showing that while exoplanet masses are limited to at most a few Jupiter masses at the onset of the GDGI instability, the companions formed can exceed the nominal gas giant mass limit of $\sim 13.5 M_{Jup}$ and form brown dwarfs as well as gas giants for the most massive disks considered. Figure 5 then depicts the companion masses as a function of initial outer disk temperature for each M-dwarf star.

Finally, Figures 6 and 7 summarize the results of the five Tables and Figures 3, 4, and 5 delineating the portions of M-dwarf star mass and disk mass space or M-dwarf star mass and disk-to-star mass ratio space that result in 100\%, greater than 50\%, less than 50\%, or 0\% frequency of gas giant planet formation. Figures 6 and 7 represent a concise summary of the existence proof for M-dwarf gas giants formed by GDGI that is the main goal of this paper. 

\section{Discussion}

Here we compare the new results with previous theoretical work for GEMS and with observational constraints on M-dwarf disk masses and exoplanet demographics. 
 
It is important to state upfront that the present models will tend to err on the side of enhancing the possibility of gas giant planet formation by GDGI, because of the simplistic radiative cooling formalism used in the Enzo 2.6 code. Figure 2 shows that the spiral features in the model temperature distributions are not nearly as well-defined as in 3D radiative hydrodynamics models of GDGI using either radiative transfer in the diffusion approximation (e.g., Boss 2021b, see Figure 2) or the beta cooling approximation (e.g., Boss 2021a, see Figures 2 and 3). Figure 4 of Boss (2021b) shows that  with a proper treatment of radiative transfer, the gas disk temperatures at the centers of dense spiral arms can be three times higher than outside the arms, compared to the much smaller enhancements in the present models (Figure 2). While the present models compute relatively quickly, thanks to AMR, they are likely to present a more optimistic estimate of the success of the GDGI mechanism than slower running, non-AMR models with more accurate gas disk thermodynamics  (e.g., Boss 2021a,b).

\subsection{M-dwarf Gas Giant Demographics}

Recent discoveries of GEMS have extended beyond early M-dwarfs to mid and late M-dwarfs (Figure 8; Morales et al. 2019; Feng et al. 2020; Kanodia et al. 2023; Hobson et al. 2023; Kagetani et al. 2023), making it harder to reconcile observations with predictions from the core accretion mechanism (e.g., Burn et al. 2021). Utilizing planet candidates from the TESS mission, Gan et al. (2023) estimated the occurrence rate of transiting hot Jupiters to be $0.27\%$ for stars with masses from 0.45 to 0.65 $M_\odot$. Similarly, Bryant et al. (2023) searched for transiting giant planets around low mass stars with masses in the range from 0.1 to 0.7 $M_\odot$, finding an occurrence rate of about $0.194 \%$. Given that these studies utilize a sample of planet candidates (and not confirmed planets), they must be treated as upper limits on the occurrence of short-period transiting GEMS. A more robust and larger sample of these objects is currently being characterized as part of the GEMS survey and will be detailed in our upcoming manuscript (Kanodia et al., in prep.). 

Regardless, this trend has been confirmed with longer period companions from some RV surveys, which are a lot more commonplace. A six-year-long radial velocity survey of 200 late M dwarfs (0.1 to 0.3 $M_\odot$) detected no Jupiter-mass companions (Pass et al., 2023). However, results from the CARMENES Doppler survey of 326 nearby M-dwarfs, Sabotta et al. (2021) determined a frequency of about 6\% (with 3\% to 10\% error bars) for giant planets with orbital periods shorter than 1000 d for a sub-sample of 71 stars. Bryant et al. (2023) suggest as solutions either that the protoplanetary disk masses are large enough for core accretion to be successful in forming giant planets, or else that the GDGI mechanism is responsible for their formation. Clearly these giant exoplanets present a problem for formation by the classic core accretion mechanism.

Boss (2023) showed how the formation of multiple planets through the GDGI mechanism can subsequently lead to mergers, migration or ejection. The Boss (2023) models were limited to solar-mass protostars but included disks with radii of 20 au, 30 au, and 60 au, finding that the different radii disks tended to produce gas giants with similar masses and orbital separations (Figures 6 and 7, Boss 2023). The present work showcases the potential of the formation of giant planets and brown-dwarf sized objects around a range of M-dwarf protostellar disks, which can subsequently be explored through detections of Jupiters, super-Jupiters and brown-dwarfs from Gaia astrometry in DR3 and DR4 (Perryman et al. 2014; Sozzetti et al. 2014, 2023). The upcoming space microlensing survey of the {\it Roman Space Telescope} will provide additional detections and constraints on the abundances of exoplanet companions to M dwarfs (Penny et al. 2019).

%  Gan et al. (2023) estimated the occurrence rate of transiting hot Jupiters to be $0.27\%$
% for stars with masses from 0.45 to 0.65 $M_\odot$. Similarly,
% Bryant et al. (2023) searched for transiting giant planets around low mass stars with masses
% in the range from 0.1 to 0.7 $M_\odot$, finding an occurrence rate of about $0.194 \%$,
% a small but non-zero rate at odds with the predictions of the core accretion mechanism
% (e.g., Burn et al. 2021). Bryant et al. (2023) suggest as solutions either that the protoplanetary
% disk masses are large enough for core accretion to be successful in forming giant planets,
% or else that the GDGI mechanism is responsible for their formation.

%  Longer period and more massive companions are even more commonplace than hot Jupiters around low
% mass stars. E.g., the CARMENES Doppler survey of 326 nearby M-dwarfs (Sabotta et al. 2021) 
% determined a frequency of about 6\% for giant planets with orbital periods shorter than 
% 1000 d for a subsample of 71 stars. 

\subsection{Theoretical Population Synthesis Models}

The classical formation mechanism for gas giant protoplanets is core accretion (Mizuno 1980), where a rocky core forms by collisional accumulation, eventually becoming massive enough that its hydrogen-rich atmosphere is unstable to contraction, leading to a phase of runaway accretion of protoplanetary disk gas. Core accretion is slower to form a gas giant planet around an M-dwarf than a solar-mass protostar, as the lower stellar mass leads to longer orbital periods at a given distance from the star. In fact, core accretion was found to be too slow to produce gas giants around M-dwarfs prior to disk gas removal (Laughlin et al. 2004). Ida \& Lin (2005) predicted that the frequency of observable gas giants formed by core accretion would decrease dramatically for stellar masses below a solar-mass, falling to zero below 0.4 $M_\odot$. Similarly, the Next Generation Planetary Population Synthesis models of Burn et al. (2021) and Schlecker et al. (2022) found that the frequency of gas giant planets formed by core accretion dropped to zero for stellar masses below 0.5 $M_\odot$. A similar inability to form gas giants occurs in the pebble accretion scenario (Liu et al. 2019). Recently, Kessler \& Alibert (2023) have raised a new concern for the core accretion mechanism for giant planet formation, finding that when numerical models included the effects of both pebble and planetesimal accretion, rather than just one or the other, the formation of giant planets is "strongly suppressed". This suppression is associated with the heating produced by planetesimal impacts following the pebble accretion phase, resulting in a hotter protoplanet atmosphere that delays the runaway gas accretion phase until after the disk gas has dissipated. Clearly core accretion has a severe problem explaining the presence of any gas giant exoplanets orbiting late M-dwarf stars, and possibly even earlier, more massive stars.

As an alternative, Boss (2006, 2011) demonstrated the possibility of gas giant planet formation by GDGI around M-dwarf stars with masses of 0.1 and 0.5 $M_\odot$, but did not calculate enough models to attempt to undertake the population synthesis attempted here. Mercer \& Stamatellos (2020) used a smoothed particle hydrodynamics (SPH) to calculate the formation of gas giants by GDGI around M-dwarfs with masses of 0.2, 0.3 and 0.4 $M_\odot$. They found that disk-to-star mass ratios between $\sim$ 0.3 and $\sim$ 0.6 were necessary for GDGI to form gas giants. These mass ratios are considerably higher than those obtained in the present models (Figure 7), where mass ratios as low as $\sim$ 0.1 were sufficient for M-dwarfs in that stellar mass range. This is likely due to the major differences in the assumed initial conditions. Mercer \& Stamatellos (2020) started their models with disks extending from 5 au outward to as far as 120 au, where the disk was defined to have a Toomre parameter $Q = 10$, i.e., quite gravitationally stable, with $Q$ rising to values greater than 30 inside about 20 au. As a result, the Mercer \& Stamatellos (2020) disks could only fragment at distances of about 50 au, and required large disk-to-star mass ratios as a result. In comparison, the present models started with relatively cool, gravitationally unstable inner disks, resulting in robust GDGI inside 4 au (e.g., Figure 2) for much lower disk-to-star mass ratios (Figure 7). Clearly observations of M-dwarf protoplanetary disk masses and sizes are required to decide whether either of these two alternative assumptions about initial conditions are appropriate.

Haworth et al. (2020) presented the results of a large suite of SPH models of massive disks around low mass stars, intended for comparison with the results of an Atacama Large Millimeter/submillimeter Array (ALMA) survey of protoplanetary disks, which found a number of disks with spiral arms that are attributed to gravitationally unstable gas disks. Their focus was on delineating the disk-to-star mass ratio as a function of star mass that separates models that either remained axisymmetric, produced spiral arms, or fragmented into clumps. For the 50 au radius disks with no stellar irradiation that are the closest to the disk models in the present study, their Figure 6 shows that as the star mass decreases from 0.5 $M_\odot$ to 0.1 $M_\odot$, the disk-to-star mass ratio needed for spiral arm formation or fragmentation increases from $\sim$ 0.1 to $\sim$ 0.25, which compares reasonably well with the results of the present models seen in Figure 7, where the disk-to-star mass ratio needed for fragmentation increases from $\sim$ 0.05 to $\sim$ 0.2 over that same star mass range. The Haworth et al. (2020) critical disk-to-star mass ratios thus agree well with the present models without stellar irradiation; however, when stellar irradiation is included, the critical disk-to-star mass ratios increase significantly, implying that the lack of stellar irradiation in the present models may be erring on the side of encouraging fragmentation. However, in the present models, fragmentation occurs at an orbital radius around 2 au, a region with a high optical depth that is impervious to direct (radially) or indirect (surface) stellar irradiation, whereas in the Haworth et al. (2020) models, fragmentation occurs at orbital radii around 50 au (their Figure 7), where the disk is expected to be optically thin and thus more susceptible to surface heating by stellar irradiation.

\subsection{M-dwarf Disk Masses}

 Bate (2018) published a population synthesis of protostellar disk masses based on his 3D SPH model of molecular cloud collapse and fragmentation, where 183 protostars formed, with a median mass of 0.21 $M_\odot$. Their disks ranged in radius from $\approx$ 10 to $\approx$ 200 au, with disk-to-star mass ratios initially typically between 0.1 and 1, and declining after $10^4$ yr. These models suggest that from a theoretical perspective, M-dwarfs might form with disk-to-star mass ratios in the range considered in the present models. 
 
Schib et al. (2021) performed a population synthesis model of the formation of protoplanetary disks, using a one-dimensional viscous accretion disk model, and found that while fragmentation by GDGI was commonplace for solar-mass protostars, fragmentation decreased significantly as the final stellar mass decreased to $\sim 0.1 M_\odot$. Schib et al. (2023) found that disk fragmentation depends critically on the molecular cloud collapse process that leads to protostellar and protoplanetary disk formation, and required the formation of large-scale disks early in the collapse phase. These studies suggest that GDGI should only be able to produce gas giants around a minor fraction of M dwarfs. However, it should be noted that the studies by Schib et al. (2021, 2023) relied on the disk achieving a Toomre $Q = 1$ in order to undergo fragmentation. In the present models, the disks that delineate the fragmenting from non-fragmenting disk masses in Figures 6 and 7 start off with Toomre $Q$ profiles that decrease with radius (cf., Figure 2 of Boss 2017) and have minimum values in the range of 1.4 to 1.6 (e.g., Boss 2006), significantly higher than unity. Hence the Schib et al. (2021, 2023) analyses might be expected to err on the side of underestimating the tendency for GDGI to succeed, compared to the present models.

The VLA/ALMA Nascent Disk and Multiplicity (VANDAM) survey has started to characterize Class 0/I embedded protostellar disks in Perseus (Tychoniec et al. 2018) and Orion (Tobin et al. 2020). These surveys are typically unable to directly determine the protostellar mass or host spectral type with the protostar embedded too deeply for spectroscopy\footnote{Protostellar masses have been determined for a few disks where the Keplerian rotation profile has been detected such as L1527 IRS in Taurus (Tobin et al. 2012). }, and direct correlations between measured luminosity and protostellar mass complicated by accretion bursts and outflows (Dunham et al. 2014). Instead, Tychoniec et al. (2018) provide disk masses for their sample of $\sim 100$ protostellar disks spanning a range of stellar types from F to M, with a median mass (dust + gas) of about 0.075 \solmass which would be sufficient to undergo instability through the GDGI mechanism discussed in this work. 

Emsenhuber et al. (2023) studied the lifetimes of protoplanetary disks subject to both internal and external photoevaporation, finding that photoevaporation is likely to be so efficient that a nominal protoplanetary disk disappears within about 1 Myr, even for a disk with an initial mass one tenth that of its host star. Given observational evidence that typical disk lifetimes are a few Myr, Emsenhuber et al. (2023) suggest that the observations can be matched by assuming disks to be more massive and more compact than their nominal disks, while noting that making disks more massive risks making them gravitationally unstable, the situation studied in the present models.  Boyden \& Eisner (2023) compared thermochemical models of protoplanetary disks with ALMA observations of 20 disks in the Orion Nebula Cluster (ONC), finding that the disks are massive (gas masses $\ge 10^{-3} M_\odot$) and compact (radii less than 100 au), implying that external photoevaporation is just getting underway in the ONC. Their Figure 13 shows that disk mass uncertainties are large enough that nearly all of their disks could have gas masses as high as $1 M_\odot$, including five of the eight M-dwarf mass stars in their sample. Such masses are more than ample for the GDGI mechanism to proceed.

\section{Conclusions}

We have shown that the GDGI mechanism can form gas giant exoplanets rapidly in orbit around M-dwarf protostars, provided that their disk masses are of order 0.02 $M_\odot$ or higher. This result stands in stark contrast to the inability of the classic core accretion mechanism to explain the formation of gas giant exoplanets around late M-dwarf stars. Observations to date of both M-dwarf disk masses and of M-dwarf exoplanet demographics suggest that the GDGI mechanism has a role to play in explaining the formation of gas giant exoplanets. Coupled with core accretion, and combined with hybrid formation mechanisms, the framework of theoretical models of giant planet formation around M-dwarf stars is slowly emerging.

\acknowledgments

We thank the referee for making a number of suggestions that have led to the improvement of this paper. The computations were performed at the Caltech Resnick High Performance Computing Center (hpc.caltech.edu) with the support of the Carnegie Institution for Science, using the Enzo code originally developed by the Laboratory for Computational Astrophysics at the University of California San Diego and now available at https://enzo-project.org/.

\clearpage
\begin{deluxetable}{lcccc}
\tablecaption{Initial conditions for the M1 models ($0.5 M_\odot$ protostar)
with varied maximum number of AMR grid refinement levels, initial outer disk temperatures (K), 
initial disk masses ($M_\odot$), and final masses of gas giant
exoplanets ($M_{Jup}$) after 100 yr of evolution. \label{tbl-1}}
\tablewidth{0pt}
\tablehead{\colhead{Model} 
& \colhead{$N_{levels}$}
& \colhead{$T_{disk}$} 
& \colhead{$M_{disk}$} 
& \colhead{$M_{exoplanets}$}}
\startdata
     M1-5-60-50  &  5  &  60  &   0.05  &  19.5, 1.2, 0.018  \\
     M1-5-50-50  &  5  &  50  &   0.05  &  16.0, 2.7 \\  
     M1-5-40-50  &  5  &  40  &   0.05  &  20., 1.3, 0.17 \\ 
     M1-5-30-50  &  5  &  30  &   0.05  &  8.3, 5.8 \\  
     M1-5-20-50  &  5  &  20  &   0.05  &  18. \\ 
     M1-5-10-50  &  5  &  10  &   0.05  &  11., 9.8, 6.9, 0.58 \\ 
\hline
     M1-6-60-50  &  6  &  60  &   0.05  &  17.9, 3.8 \\  
     M1-6-50-50  &  6  &  50  &   0.05  &  17.5, 9.0 \\  
     M1-6-40-50  &  6  &  40  &   0.05  &  18.1, 7.8, 5.5, 0.14 \\      
     M1-6-30-50  &  6  &  30  &   0.05  &  17.9, 1.7, 1.5 \\  
     M1-6-20-50  &  6  &  20  &   0.05  &  21., 0.12 \\  
     M1-6-10-50  &  6  &  10  &   0.05  &  15., 8., 4.8, 0.29 \\ 
\hline
     M1-5-60-40  &  5  &  60  &   0.04  &  2.6, 0.68 \\                     
     M1-5-50-40  &  5  &  50  &   0.04  &  1.4, 0.33 \\                      
     M1-5-40-40  &  5  &  40  &   0.04  &  1.7 \\                          
     M1-5-30-40  &  5  &  30  &   0.04  &  3.8 \\                        
     M1-5-20-40  &  5  &  20  &   0.04  &  1.9, 0.38 \\                    
     M1-5-10-40  &  5  &  10  &   0.04  &  3.0, 0.17, 0.1 \\                      
\hline
     M1-6-60-40  &  6  &  60  &   0.04  &  4.8 \\                     
     M1-6-50-40  &  6  &  50  &   0.04  &  0.51, 0.4 \\                     
     M1-6-40-40  &  6  &  40  &   0.04  &  5.5, 2.5 \\                       
     M1-6-30-40  &  6  &  30  &   0.04  &  0.6 \\                       
     M1-6-20-40  &  6  &  20  &   0.04  &  0.61, 0.29 \\                     
     M1-6-10-40  &  6  &  10  &   0.04  &  0.4, 0.17, 0.11 \\                      
\hline
     M1-5-60-30  &  5  &  60  &   0.03  &  0.17 \\
     M1-5-50-30  &  5  &  50  &   0.03  &  0.11 \\
     M1-5-40-30  &  5  &  40  &   0.03  & 0.26 \\
     M1-5-30-30  &  5  &  30  &   0.03  & none \\  
     M1-5-20-30  &  5  &  20  &   0.03  & none \\
     M1-5-10-30  &  5  &  10  &   0.03  & 0.17 MJ \\
\hline
     M1-6-60-30  &  6  &  60  &   0.03  &  0.29 \\
     M1-6-50-30  &  6  &  50  &   0.03  &  none \\
     M1-6-40-30  &  6  &  40  &   0.03  &  0.18 \\
     M1-6-30-30  &  6  &  30  &   0.03  &  none \\  
     M1-6-20-30  &  6  &  20  &   0.03  &  none \\
     M1-6-10-30  &  6  &  10  &   0.03  &  none \\
\hline
     M1-6-60-20  &  6  &  60  &   0.02  &  none \\
     M1-6-50-20  &  6  &  50  &   0.02  &  none \\
     M1-6-40-20  &  6  &  40  &   0.02  &  none \\  
     M1-6-30-20  &  6  &  30  &   0.02  &  none \\   
     M1-6-20-20  &  6  &  20  &   0.02  &  none \\ 
     M1-6-10-20  &  6  &  10  &   0.02  &  none \\
\enddata
\end{deluxetable}

\begin{deluxetable}{lcccc}
\tablecaption{Initial conditions for the M2.5 models ($0.4 M_\odot$ protostar)
with varied maximum number of AMR grid refinement levels, initial outer disk temperatures (K), 
initial disk masses ($M_\odot$), and final masses of gas giant
exoplanets ($M_{Jup}$) after 100 yr of evolution. \label{tbl-2}}
\tablewidth{0pt}
\tablehead{\colhead{Model} 
& \colhead{$N_{levels}$}
& \colhead{$T_{disk}$} 
& \colhead{$M_{disk}$} 
& \colhead{$M_{exoplanets}$}}
\startdata
     M2.5-6-60-50 &   6 &   60 &    0.05 & 18., 6.7, 3.8 \\           
     M2.5-6-50-50 &   6 &   50 &    0.05 & 11.5, 9.7, 6.6, 0.35 \\   
     M2.5-6-40-50 &   6 &   40 &    0.05 & 20., 8.1, 1.7 \\        
     M2.5-6-30-50 &   6 &   30 &    0.05 & 19.5, 16. \\    
     M2.5-6-20-50 &   6 &   20 &    0.05 & 22., 3.1, 0.85 \\         
     M2.5-6-10-50 &   6 &   10 &    0.05 & 22.5, 1.8, 0.54 \\    
\hline
     M2.5-6-60-40 &   6 &   60 &    0.04 & 9.0, 0.88 \\   
     M2.5-6-50-40 &   6 &   50 &    0.04 & 6.1, 6.7 \\            
     M2.5-6-40-40 &   6 &   40 &    0.04 & 10.5, 1.6, 0.44 \\        
     M2.5-6-30-40 &   6 &   30 &    0.04 & 0.51, 1.4, 10. \\         
     M2.5-6-20-40 &   6 &   20 &    0.04 & 1.9, 9.5. 0.17 \\          
     M2.5-6-10-40 &   6 &   10 &    0.04 & 9.6 \\                 
\hline
     M2.5-5-60-30 &   5 &   60 &    0.03 & none \\                  
     M2.5-5-50-30 &   5 &   50 &    0.03 & 0.31 \\                  
     M2.5-5-40-30 &   5 &   40 &    0.03 & 0.20, 0.15 \\           
     M2.5-5-30-30 &   5 &   30 &    0.03 & none \\                  
     M2.5-5-20-30 &   5 &   20 &    0.03 & none \\                  
     M2.5-5-10-30 &   5 &   10 &    0.03 & 0.21 \\              
\hline
     M2.5-6-60-30 &   6 &   60 &    0.03 & none \\                 
     M2.5-6-50-30 &   6 &   50 &    0.03 & none \\                
     M2.5-6-40-30 &   6 &   40 &    0.03 & none \\              
     M2.5-6-30-30 &   6 &   30 &    0.03 & none \\                 
     M2.5-6-20-30 &   6 &   20 &    0.03 & none \\                
     M2.5-6-10-30 &   6 &   10 &    0.03 & none \\              
\hline
     M2.5-6-60-20 &   6 &   60 &    0.02 & none \\                 
     M2.5-6-50-20 &   6 &   50 &    0.02 & none \\                    
     M2.5-6-40-20 &   6 &   40 &    0.02 & none \\                   
     M2.5-6-30-20 &   6 &   30 &    0.02 & none \\                 
     M2.5-6-20-20 &   6 &   20 &    0.02 & none \\                
     M2.5-6-10-20 &   6 &   10 &    0.02 & none \\                
\enddata
\end{deluxetable}

\begin{deluxetable}{lcccc}
\tablecaption{Initial conditions for the M3.5 models ($0.3 M_\odot$ protostar)
with varied maximum number of AMR grid refinement levels, initial outer disk temperatures (K), 
initial disk masses ($M_\odot$), and final masses of gas giant
exoplanets ($M_{Jup}$) after 100 yr of evolution. \label{tbl-3}}
\tablewidth{0pt}
\tablehead{\colhead{Model} 
& \colhead{$N_{levels}$}
& \colhead{$T_{disk}$} 
& \colhead{$M_{disk}$} 
& \colhead{$M_{exoplanets}$}}
\startdata
     M3.5-6-60-40 &  6 &   60 &    0.04   &  13., 6.8, 3.7            \\
     M3.5-6-50-40 &  6 &   50 &    0.04   &  14., 4.3                 \\
     M3.5-6-40-40 &  6 &   40 &    0.04   &  8.2, 1.7, 1.5, 1.4, 0.8  \\
     M3.5-6-30-40 &  6 &   30 &    0.04   &  11., 7.8, 1.3            \\
     M3.5-6-20-40 &  6 &   20 &    0.04   &  9.7, 5.8, 4.8            \\
     M3.5-6-10-40 &  6 &   10 &    0.04   &  8.9, 6.3, 5.0            \\
\hline
     M3.5-5-60-40 &  5 &   60 &    0.04   &  10.5, 0.5, 0.4           \\ 
     M3.5-5-50-40 &  5 &   50 &    0.04   &  13.                      \\
     M3.5-5-40-40 &  5 &   40 &    0.04   &  16.                      \\
     M3.5-5-30-40 &  5 &   30 &    0.04   &  4.9, 2.1                 \\
     M3.5-5-20-40 &  5 &   20 &    0.04   &  9.95                     \\
     M3.5-5-10-40 &  5 &   10 &    0.04   &  12.                      \\
\hline
     M3.5-6-60-30 &  6 &   60 &    0.03   &  none                    \\
     M3.5-6-50-30 &  6 &   50 &    0.03   &  0.11                     \\
     M3.5-6-40-30 &  6 &   40 &    0.03   &  0.10                     \\
     M3.5-6-30-30 &  6 &   30 &    0.03   &  none                     \\
     M3.5-6-20-30 &  6 &   20 &    0.03   &  none                     \\
     M3.5-6-10-30 &  6 &   10 &    0.03   &  3.4                      \\
\hline
     M3.5-5-60-30 &  5 &   60 &    0.03   &  0.15                     \\
     M3.5-5-50-30 &  5 &   50 &    0.03   &  0.1                      \\
     M3.5-5-40-30 &  5 &   40 &    0.03   &  0.4                      \\
     M3.5-5-30-30 &  5 &   30 &    0.03   &  none                     \\
     M3.5-5-20-30 &  5 &   20 &    0.03   &  0.35                     \\
     M3.5-5-10-30 &  5 &   10 &    0.03   &  0.8                      \\
\hline
     M3.5-6-60-20 &  6 &   60 &    0.02   &  none                     \\
     M3.5-6-50-20 &  6 &   50 &    0.02   &  none                     \\
     M3.5-6-40-20 &  6 &   40 &    0.02   &  none                     \\
     M3.5-6-30-20 &  6 &   30 &    0.02   &  none                     \\
     M3.5-6-20-20 &  6 &   20 &    0.02   &  none                     \\
     M3.5-6-10-20 &  6 &   10 &    0.02   &  none                     \\
\enddata
\end{deluxetable}

\begin{deluxetable}{lcccc}
\tablecaption{Initial conditions for the M4.5 models ($0.2 M_\odot$ protostar)
with varied maximum number of AMR grid refinement levels, initial outer disk temperatures (K), 
initial disk masses ($M_\odot$), and final masses of gas giant
exoplanets ($M_{Jup}$) after 100 yr of evolution. \label{tbl-4}}
\tablewidth{0pt}
\tablehead{\colhead{Model} 
& \colhead{$N_{levels}$}
& \colhead{$T_{disk}$} 
& \colhead{$M_{disk}$} 
& \colhead{$M_{exoplanets}$}}
\startdata
     M4.5-6-60-40 &  6 &   60 &   0.04 & 14., 3.2, 3.0, 2.5\\      
     M4.5-6-50-40 &  6 &   50 &   0.04 & 16.5, 6.7, 0.21       \\   
     M4.5-6-40-40 &  6 &   40 &   0.04 & 13.5, 6.5, 5.2, 2.5  \\
     M4.5-6-30-40 &  6 &   30 &   0.04 & 11.5, 7.8, 4.7, 4.1 \\     
     M4.5-6-20-40 &  6 &   20 &   0.04 & 18.5, 1.35, 0.86, 0.43 \\
     M4.5-6-10-40 &  6 &   10 &   0.04 & 18., 3.4, 1.8, 0.18 \\      
\hline
     M4.5-5-60-30 &  5 &   60 &   0.03 &    0.44 \\             
     M4.5-5-50-30 &  5 &   50 &   0.03 &    0.26 \\                      
     M4.5-5-40-30 &  5 &   40 &   0.03 &    0.84 \\                     
     M4.5-5-30-30 &  5 &   30 &   0.03 &    0.25 \\                  
     M4.5-5-20-30 &  5 &   20 &   0.03 &    0.5 \\                  
     M4.5-5-10-30 &  5 &   10 &   0.03 &    0.45, 1.1 \\             
\hline
     M4.5-6-60-30 &  6 &   60 &   0.03 &    none \\                  
     M4.5-6-50-30 &  6 &   50 &   0.03 &    0.11 \\                      
     M4.5-6-40-30 &  6 &   40 &   0.03 &    0.27 \\                     
     M4.5-6-30-30 &  6 &   30 &   0.03 &    0.12 \\                  
     M4.5-6-20-30 &  6 &   20 &   0.03 &    0.21 \\                  
     M4.5-6-10-30 &  6 &   10 &   0.03 &    0.11 \\                  
\hline
     M4.5-6-60-20 &  6 &   60 &   0.02 &    none  \\                  
     M4.5-6-50-20 &  6 &   50 &   0.02 &    none  \\                
     M4.5-6-40-20 &  6 &   40 &   0.02 &    none  \\              
     M4.5-6-30-20 &  6 &   30 &   0.02 &    none  \\            
     M4.5-6-20-20 &  6 &   20 &   0.02 &    none  \\          
     M4.5-6-10-20 &  6 &   10 &   0.02 &    none  \\        
\hline
     M4.5-5-30-20 &  5 &   30 &   0.02 &    none  \\      
     M4.5-5-20-20 &  5 &   20 &   0.02 &    none  \\                     
     M4.5-5-10-20 &  5 &   10 &   0.02 &    0.25  \\                 
\hline
     M4.5-6-30-15 &  6 &   30 &   0.015 &   none \\                 
     M4.5-6-20-15 &  6 &   20 &   0.015 &   none \\                  
     M4.5-6-10-15 &  6 &   10 &   0.015 &   none \\                  
\hline
     M4.5-5-30-15 &  5 &   30 &   0.015 &   none \\                    
     M4.5-5-20-15 &  5 &   20 &   0.015 &   none \\                  
     M4.5-5-10-15 &  5 &   10 &   0.015 &   none \\                    
\hline
     M4.5-6-30-10 &  6 &   30 &   0.01 &   none \\                 
     M4.5-6-20-10 &  6 &   20 &   0.01 &   none \\                  
     M4.5-6-10-10 &  6 &   10 &   0.01 &   none \\                  
\enddata
\end{deluxetable}
\clearpage

\begin{deluxetable}{lcccc}
\tablecaption{Initial conditions for the M6 models ($0.1 M_\odot$ protostar)
with varied maximum number of AMR grid refinement levels, initial outer disk temperatures (K), 
initial disk masses ($M_\odot$), and final masses of gas giant
exoplanets ($M_{Jup}$) after 100 yr of evolution. \label{tbl-5}}
\tablewidth{0pt}
\tablehead{\colhead{Model} 
& \colhead{$N_{levels}$}
& \colhead{$T_{disk}$} 
& \colhead{$M_{disk}$} 
& \colhead{$M_{exoplanets}$}}
\startdata
     M6-6-30-30  &  6  &  30    &   0.030 &  11., 1.1, 0.95, 3.6 \\     
     M6-6-20-30  &  6  &  20    &  0.030 &  11., 9.2, 0.22      \\     
     M6-6-10-30  &  6  &  10    &   0.030 &  8.3, 4.4, 2.4, 2.1  \\      
\hline
     M6-5-30-25  &  5  &  30    &   0.025 &   0.2, 0.16 \\        
     M6-5-20-25  &  5  &  20    &   0.025 &   0.25 \\                
     M6-5-10-25  &  5  &  10    &   0.025 &   4. \\                     
\hline
     M6-6-30-25  &  6  &  30    &   0.025 &   0.21 \\                 
     M6-6-20-25  &  6  &  20    &   0.025 &   none \\                   
     M6-6-10-25  &  6  &  10    &   0.025 &   3.2, 0.2 \\           
\hline
     M6-4-30-20  &  4  &  30    &   0.020 &   none \\                 
     M6-4-20-20  &  4  &  20    &   0.020 &   none \\                
     M6-4-10-20  &  4  &  10    &   0.020 &   0.3, 0.2 \\           
\hline
     M6-5-30-20  &  5  &  30    &   0.020 &   none  \\                  
     M6-5-20-20  &  5  &  20    &   0.020 &   0.1 \\                  
     M6-5-10-20  &  5  &  10    &   0.020 &   0.2 \\                  
\hline
     M6-6-30-20  &  5  &  30    &   0.020 &   none \\                  
     M6-6-20-20  &  5  &  20    &   0.020 &   none \\                                
     M6-6-10-20  &  5  &  10    &   0.020 &   none \\                 
\hline
     M6-5-30-15  &  5  &  30    &  0.015 &   none \\                 
     M6-5-20-15  &  5  &  20    &   0.015 &   none \\                   
     M6-5-10-15  &  5  &  10    &   0.015 &   none \\                 
\hline
     M6-5-30-10  &  5  &  30    &   0.010 &   none \\                        
     M6-5-20-10  &  5  &  20    &  0.010 &   none \\                     
     M6-5-10-10  &  5  &  10    &  0.010 &   none \\                     
\enddata
\end{deluxetable}
\clearpage

% \begin{figure}
% \vspace{-1.0in}
% \includegraphics[scale=.60,angle=-90]{f1.pdf}
% \vspace{0.5in}
% \caption{Initial log density cross-section in a vertical section ($x$ = 0) showing the entire computational 
% grid with a maximum of six levels of refinement for model M1-6-40-40, a 0.04 $M_\odot$ disk in orbit 
% around a 0.5 $M_\odot$ protostar.}\label{fig:1}
% \end{figure}
% \clearpage

% \begin{figure}
% \vspace{-1.0in}
% \includegraphics[scale=.60,angle=-90]{f2.pdf}
% \vspace{0.5in}
% \caption{Initial log density cross-section in the disk midplane ($z$ = 0) showing the entire computational 
% grid with a maximum of six levels of refinement for model M1-6-40-40, a 0.04 $M_\odot$ disk in orbit 
% around a 0.5 $M_\odot$ protostar.}\label{fig:2}
% \end{figure}

\begin{figure*}[ht]
\centering
\begin{tabular}{cc}
\hspace{-2 cm}
 \includegraphics[width=8 cm,angle=-90]{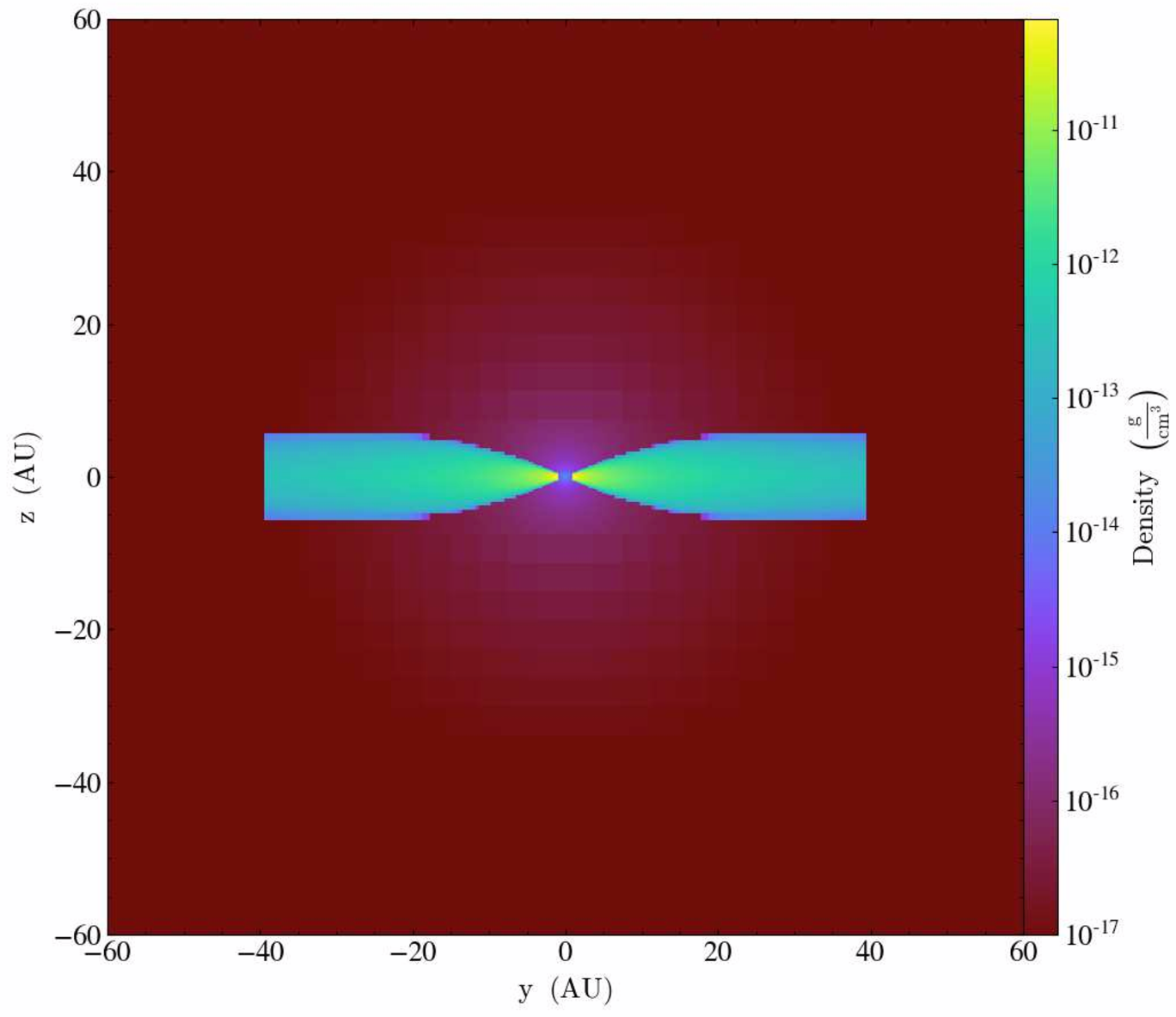} & \hspace{-1 cm}
 \includegraphics[width=8 cm,angle=-90]{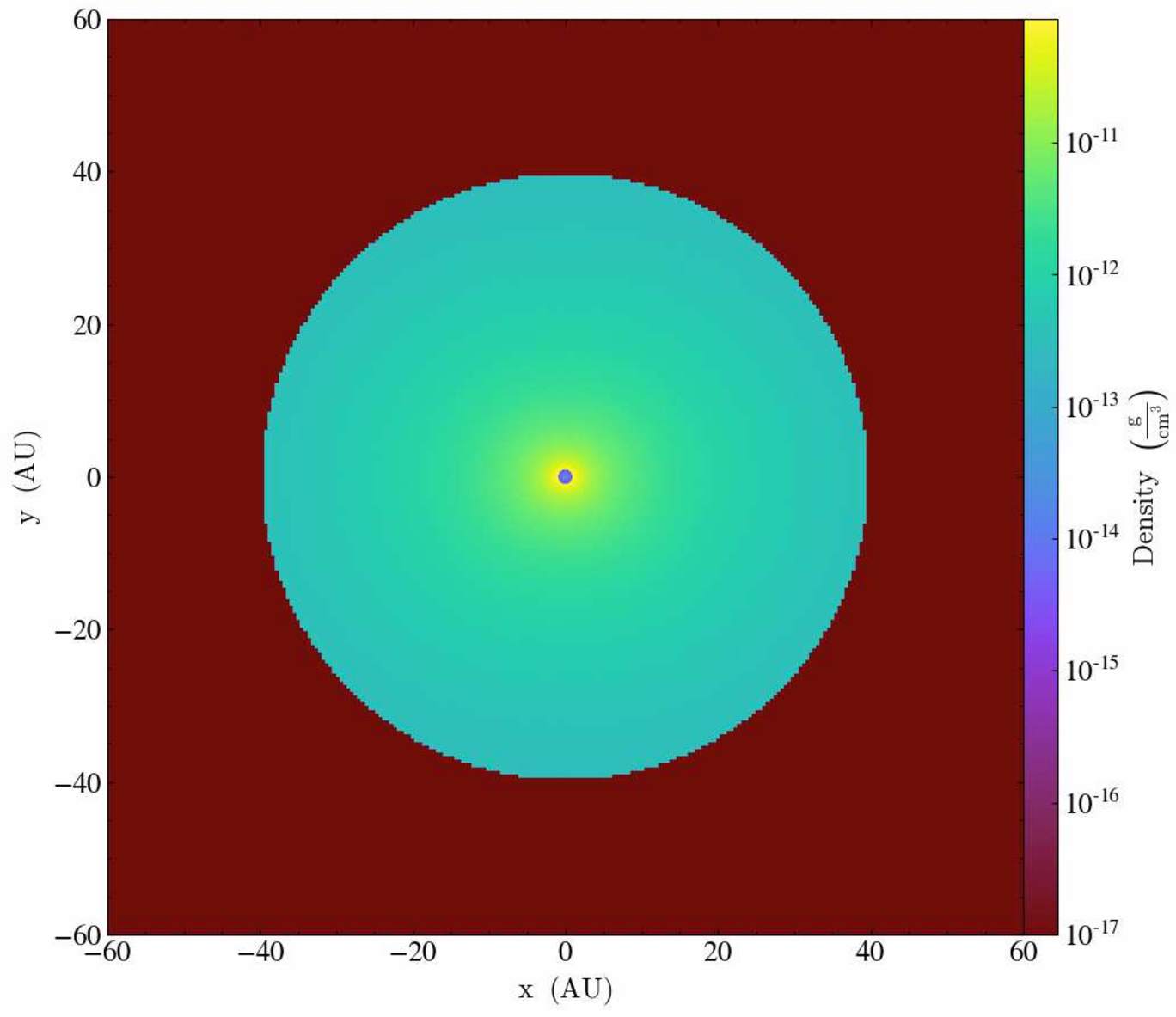} \\
\end{tabular}
\caption{Initial log density cross-section in a vertical section ($x$ = 0) on the left, and in the disk midplane ($z$ = 0) on the right, showing the entire computational 
grid with a maximum of six levels of refinement for model M1-6-40-40, a 0.04 $M_\odot$ disk in orbit 
around a 0.5 $M_\odot$ protostar.}   \label{fig:initialcrosssection}
\end{figure*}
\clearpage

\begin{figure}
\vspace{-2.0in}
\centering
\includegraphics[scale=0.6,angle=90]{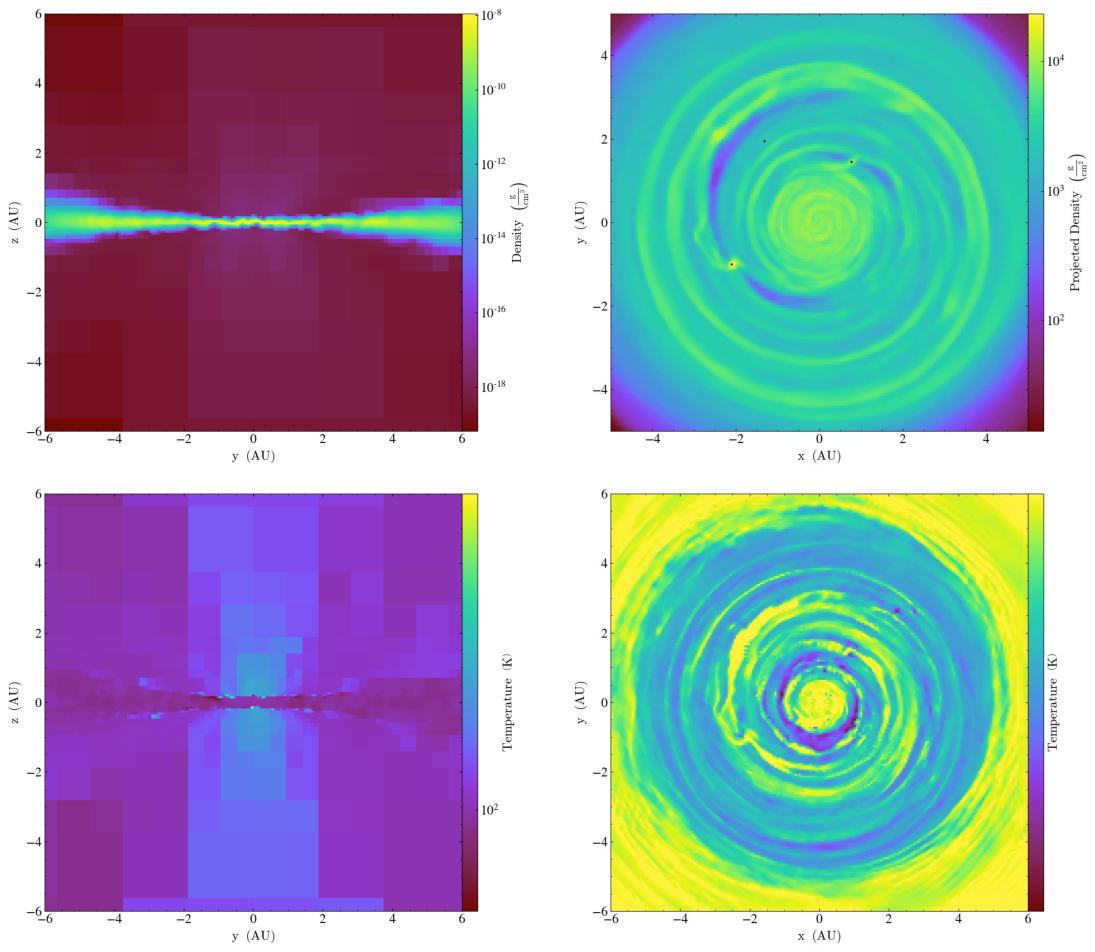}
%\includegraphics[width=8in,trim={6cm 1cm 2cm 0cm},angle=90]{FinalCrossSection.ps}
% \vspace{0.5in}
\caption{Cross-sections after 100 years of evolution for model M1-6-40-40. \textbf{Top Left: } Log density cross-section along the rotation axis for the plane $x$ = 0. \textbf{Bottom Left:} Temperature cross-section along the rotation axis for the plane $x$ = 0. Temperatures range from about 85 K (dark blue) at the disk midplane to about
240 K (yellow) in the regions above and below the disk. \textbf{Top Right:} Log density cross-section in the disk midplane ($z$ = 0) showing two gas giant protoplanets in the inner disk, one at 1 o'clock with a
mass of 2.5 $M_{Jup}$ and one at 8 o'clock with a mass of 5.5 $M_{Jup}$. The sink particle
at 11 o'clock has a mass less than 0.1 $M_{Jup}$ and so is not tabulated. Dense clumps
which have not formed sink particles can be seen as well, e.g., at 10 o'clock. \textbf{Bottom Right:} Temperature cross-section in the disk midplane ($z$ = 0). Midplane temperatures range from about
85 K (dark blue) to about 100 K (yellow).}\label{fig:finalcrosssection}
\end{figure}

\clearpage

\begin{figure}
\vspace{-0.5in}
\includegraphics[scale=0.7,angle=+90]{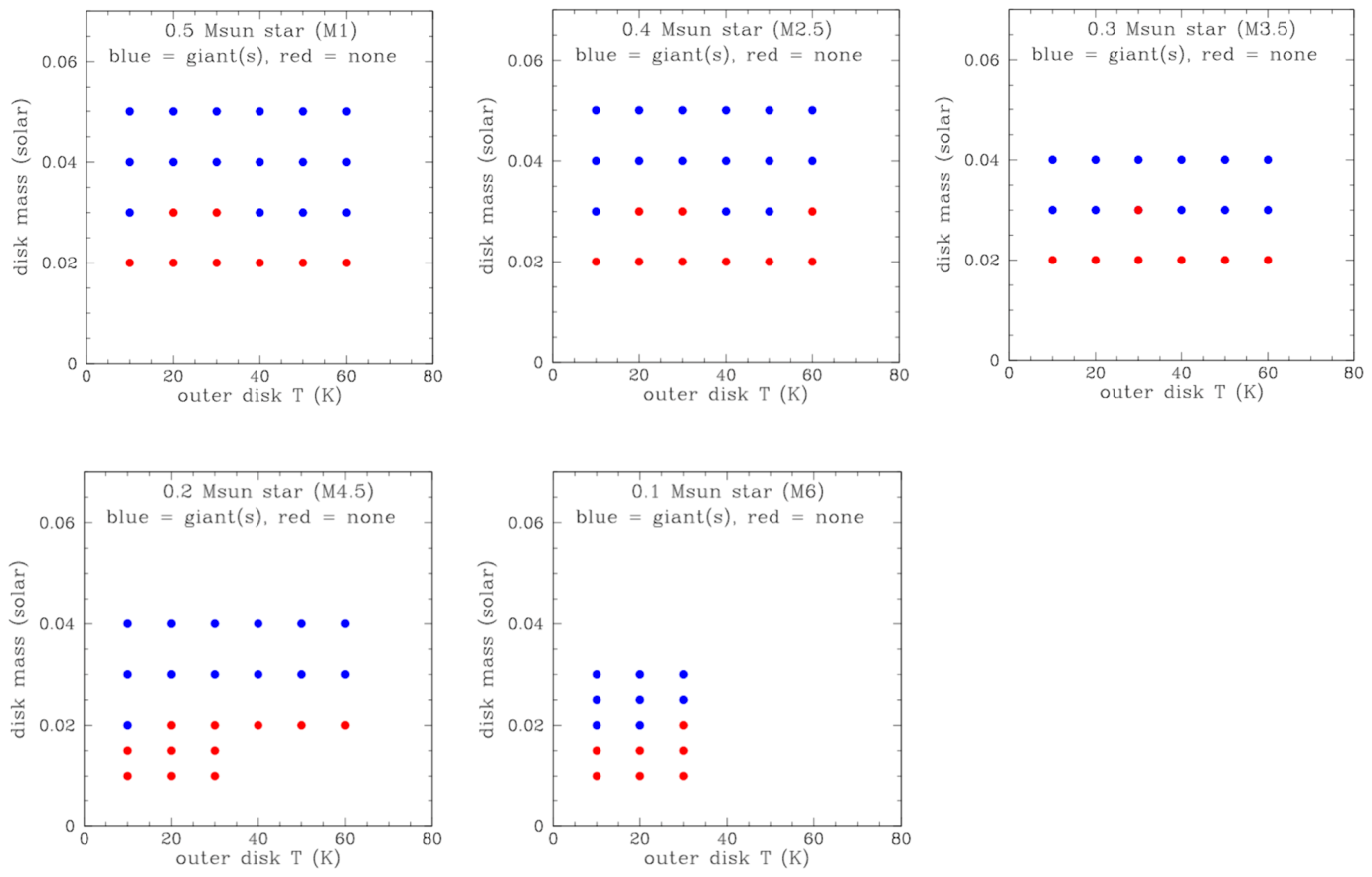}
\vspace{-0.5in}
\caption{Dependence of gas giant protoplanet formation by GDGI on M-dwarf stellar mass,
disk mass, and outer disk temperatures for the 162 models in Tables 1 through 5.}\label{fig:7}
\end{figure}
\clearpage

\begin{figure}
\vspace{-0.5in}
\includegraphics[scale=0.7,angle=+90]{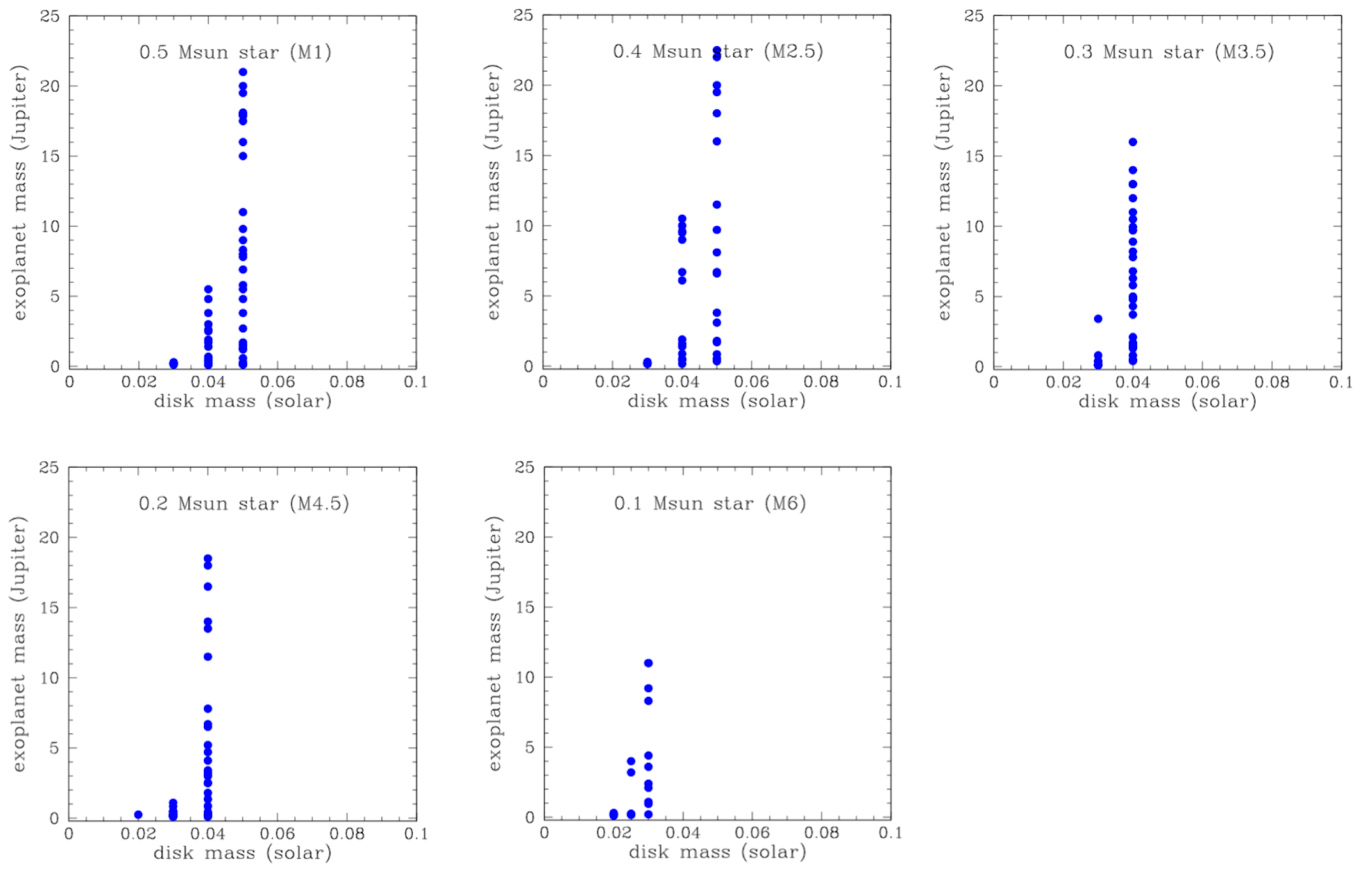}
\vspace{-0.5in}
\caption{Gas giant exoplanet masses as a function of the disk masses for the 162 models in
Tables 1 through 5. }\label{fig:8}
\end{figure}
\clearpage

\begin{figure}
\vspace{-0.5in}
\includegraphics[scale=0.7,angle=+90]{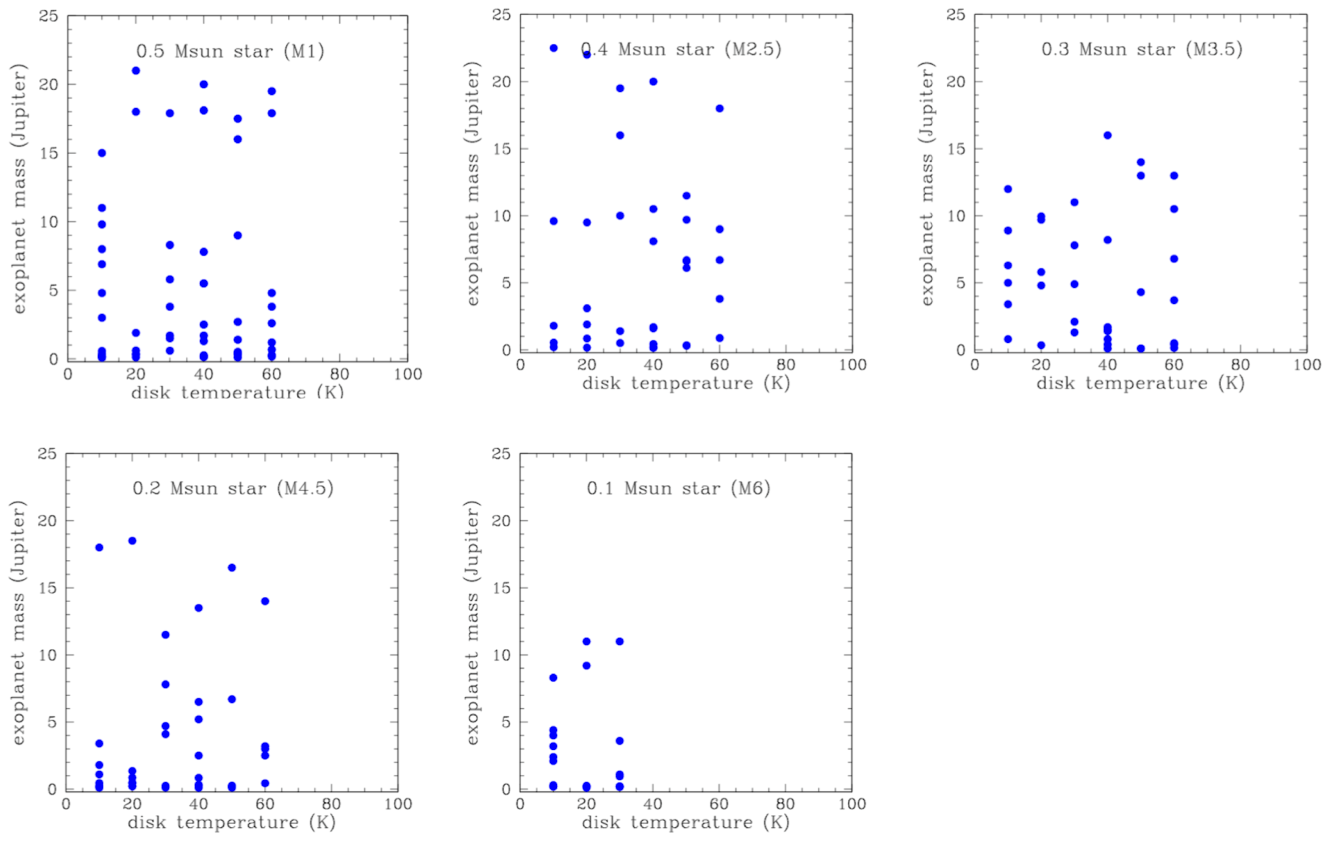}
\vspace{-0.5in}
\caption{ Gas giant exoplanet masses as a function of the outer disk temperatures for the 162 models in
Tables 1 through 5.  }\label{fig:9}
\end{figure}
\clearpage

% \begin{figure}
% \vspace{-2.0in}
% \includegraphics[scale=0.8,angle=-90]{f10.pdf}
% \vspace{+0.5in}
% \caption{Gas giant exoplanet formation frequency by GDGI as a function of M star mass
% and protoplanetary disk mass. Disk masses of $\sim 0.02 M_\odot$ or higher appear to
% be required for the GDGI to produce gas giant protoplanets.}\label{fig:10}
% \end{figure}
% \clearpage

\begin{figure}
\vspace{-2.0in}
\includegraphics[scale=0.6,angle=90]{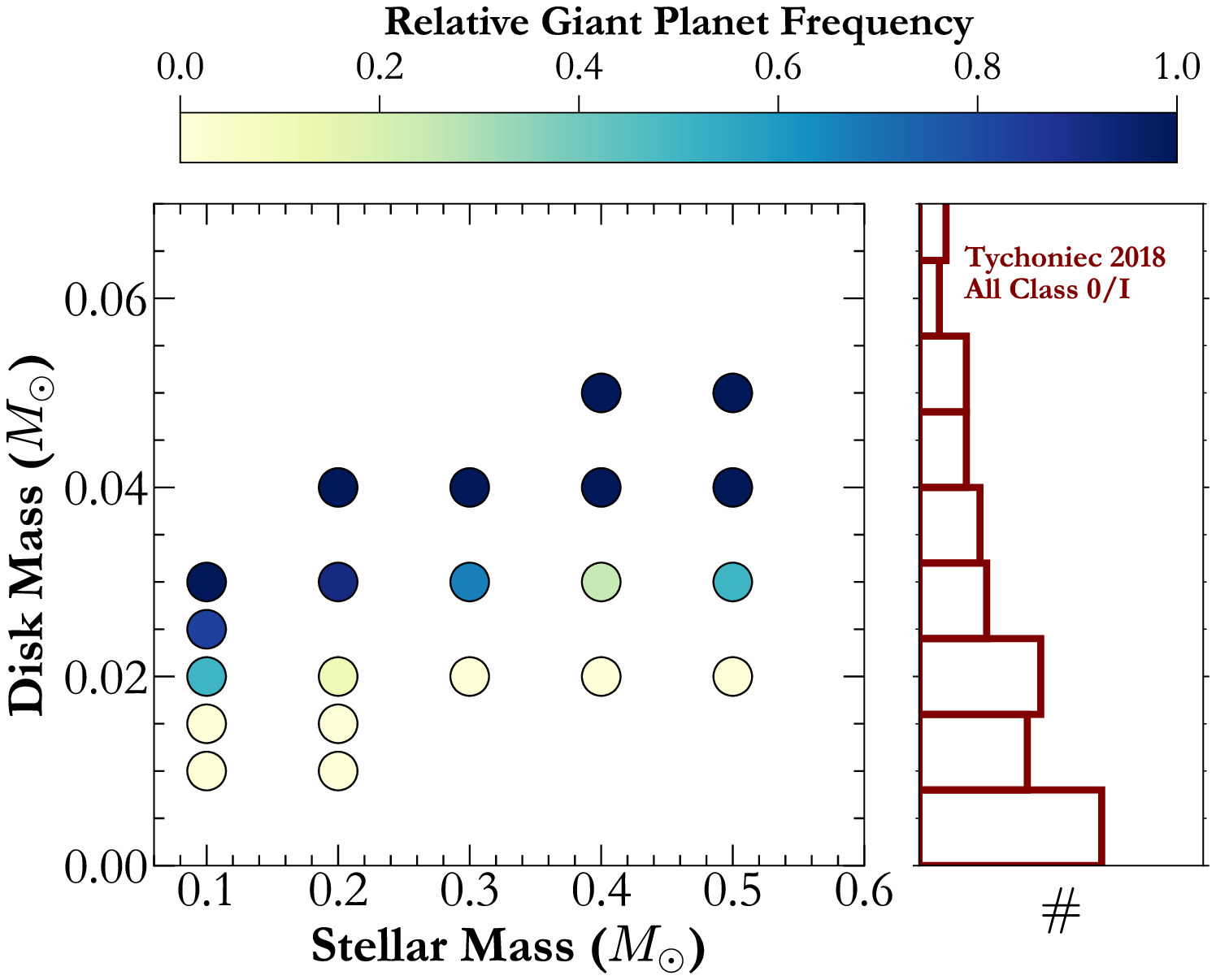}
% \vspace{+0.5in}
\caption{Gas giant exoplanet formation frequency by GDGI as a function of M star mass and protoplanetary disk mass. Disk masses of $\sim 0.02 M_\odot$ or higher appear to be required for the GDGI to produce gas giant protoplanets. The histogram on the right shows the distribution of Class 0/I disk masses (gas + dust) as measured by VLA for $\sim$ 100 disks in Perseus (Tychoniec et al. 2018). We note the caveat that this sample consists of embedded protostellar FGKM sources, and is therefore likely biased to have a higher median disk mass than an M-dwarf only sample. However, given the difficulty in estimating host stellar masses for embedded protostellar disks, such an M-dwarf only protostellar sample does not currently exist.}\label{fig:10}
\end{figure}
\clearpage

\begin{figure}
\vspace{-2.0in}
\includegraphics[scale=0.6,angle=90]{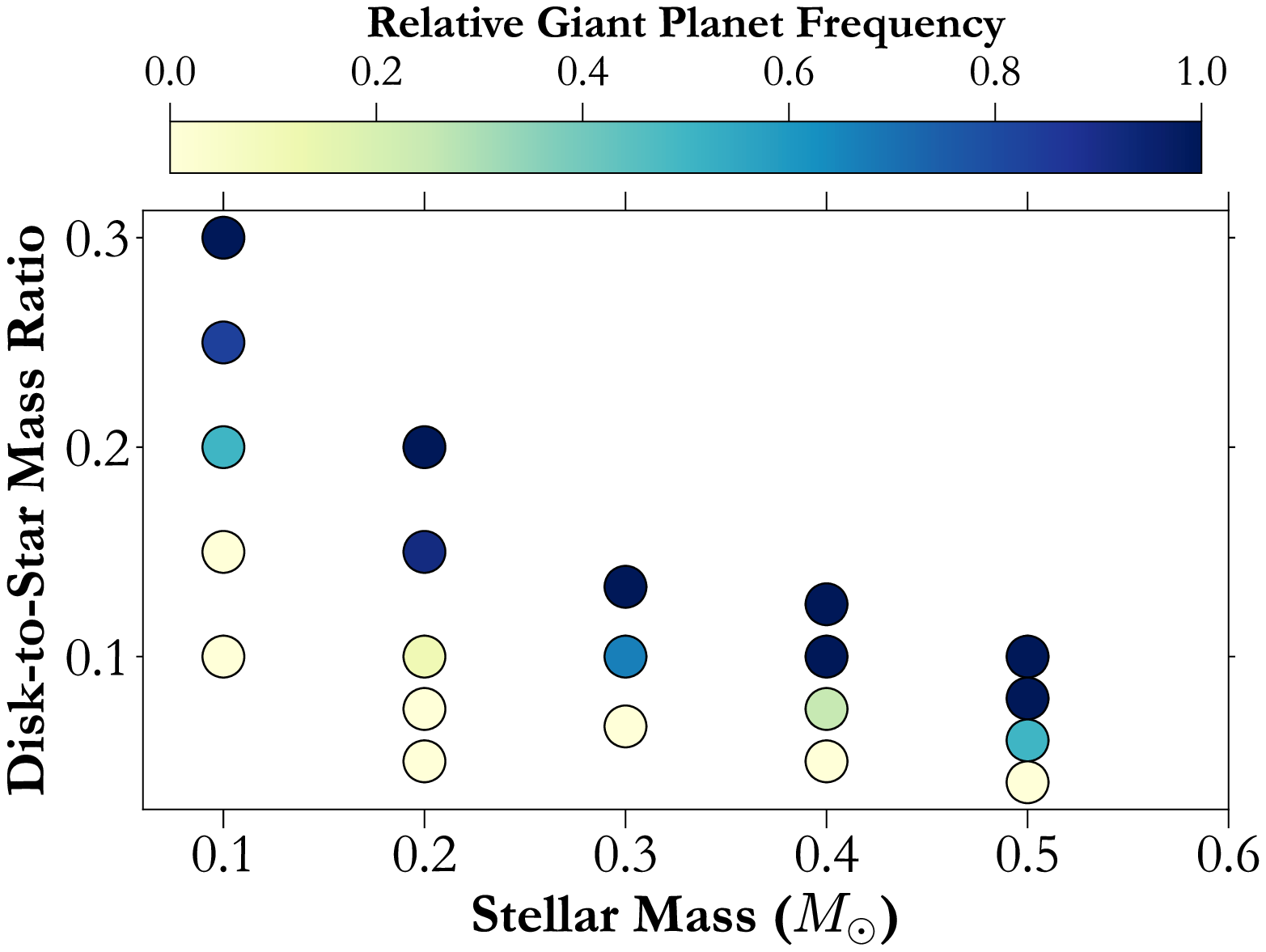}
\vspace{+0.5in}
\caption{Gas giant exoplanet formation frequency by GDGI as a function of M star mass
and protoplanetary disk-to-star mass ratio.}\label{fig:11}
\end{figure}
\clearpage

\begin{figure}
\vspace{-2.0in}
\includegraphics[scale=0.6,angle=90]{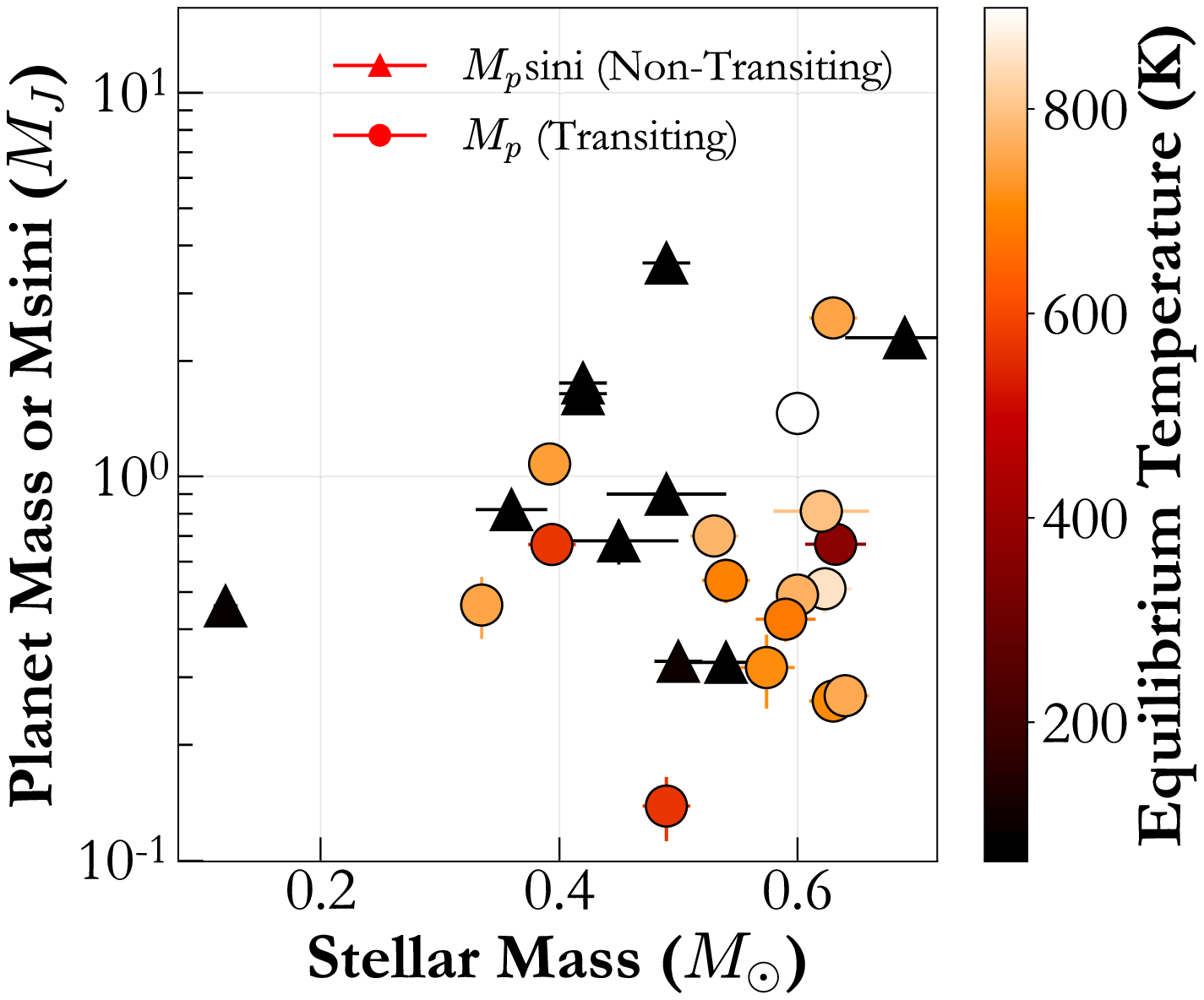}
\vspace{+0.5in}
\caption{Planetary mass plotted as a function of stellar mass for GEMS. The transiting planets have true mass measurements, whereas \msini{} is plotted for the non-transiting ones, with the exception of GJ 463~b, which has a true mass measurement of $\sim 1140$ \earthmass{} from astrometry (Sozzetti et al. 2023).}\label{fig:sample}
\end{figure}
\clearpage

\end{document}